\documentclass[reprint,showpacs,preprintnumbers,nofootinbib,amsmath,amssymb,aps,prd,floatfix]{revtex4-1}
\date{\today}
\begin{document}



\title{Mixed states for mixing neutrinos}



\author{Gabriel Cozzella}\email{gabriel.cozzella@unesp.br}    
\affiliation{Instituto de F\'\i sica Te\'orica, 
Universidade Estadual Paulista, Rua Dr.\ Bento Teobaldo Ferraz, 271, 01140-070, S\~ao Paulo, S\~ao Paulo, Brazil}

\author{Carlo Giunti}\email{carlo.giunti@to.infn.it}
\affiliation{INFN, Sezione di Torino, Via P. Giuria 1, I--10125 Torino, Italy}



\begin{abstract}
Here we discuss the description of flavor neutrinos produced
or detected in processes which involve more than one neutrino.
We show that in these cases flavor neutrinos cannot be separately described by pure states,
but require a density matrix description.
We consider explicitly the examples of $\nu_{e}$ and $\bar\nu_{\mu}$ production in
$\mu^{+}$ decay
and
$\nu_{\mu}$ detection
through scattering on electrons.
We show that
the density matrix which describes a flavor neutrino
can be approximated with a density matrix of a pure state only when the differences of the neutrino masses are neglected
in the interaction process.
In this approximation,
the pure states are the standard flavor states and one recovers the standard expression
for the neutrino oscillation probability.
We discuss also the effects of mixing of the three standard light neutrinos
with heavy neutrinos which can be either decoupled
because their masses are much larger than the
maximum neutrino energy in the neutrino production process
or because they are produced and detected incoherently.
Finally, we discuss
the more complicated case of neutrino-electron elastic scattering,
in which the initial and final neutrinos do not have determined flavors,
but there is a flavor dependence due to the different contributions of
charged-current and neutral-current interactions.
\end{abstract}


\maketitle

\section{Introduction}\label{sec:I}

In the last years neutrino physics has proven to be a fertile ground for particle theory.
The discovery \cite{Fukuda:1998mi,Ahmad:2002jz}
of neutrino oscillations
\cite{Pontecorvo:1957cp,Pontecorvo:1957qd,Maki:1962mu}
has been awarded the 2015 Nobel prize, recognizing its importance.
Neutrino oscillations imply that neutrinos must have non-zero masses
and that there is neutrino mixing.
The standard theory of neutrino mixing and oscillations is well known
(see, for example, Refs.~\cite{Giunti:2007ry,Bilenky:2018hbz,Patrignani:2016xqp}),
but there are subtle issues that require a special treatment
(see, for example, the recent discussions in Refs.~\cite{Akhmedov:2010ua,Akhmedov:2012uu,Akhmedov:2017xxm}).

Neutrino oscillations are transitions among different neutrino flavors
that can be observed at macroscopic distances from a neutrino source.
Different neutrino flavors
($\nu_{e}$, $\nu_{\mu}$, $\nu_{\tau}$)
are characterized by their production or detection
in association with the corresponding charged lepton
($e$, $\mu$, $\tau$).
In the standard treatment of neutrino oscillations,
flavor neutrinos are described by states which are
unitary superpositions of massive neutrino states
and the mixing matrix is the unitary matrix that
diagonalizes the mass matrix of the neutrino fields
(see Refs.~\cite{Giunti:2007ry,Bilenky:2018hbz}).
However, in the description of neutrinos as excitations of quantum fields, the relation between mass and flavor states is not as simple, due to the non-existence of a canonical set of creation and annihilation operators for the flavor fields \cite{Giunti:1991cb}.
This fact implies that
the neutrino flavor states are phenomenological quantities that describe neutrinos
created or detected in a weak interaction process
as superpositions of massive neutrinos with coefficients determined
by the respective interaction amplitudes
\cite{Giunti:1991cb,Bilenky:2001yh,Giunti:2002xg,Giunti:2006fr}.
The standard neutrino flavor states are recovered in the realistic approximation
of neglecting the neutrino mass differences in the interaction process.

The localization of the production and detection processes
in a neutrino oscillation experiment
and the associated energy-momentum uncertainties
require a wave-packet description \cite{Kayser:1981ye}
(see Refs.~\cite{Beuthe:2001rc,Giunti:2003ax,Giunti:2007ry}).
It has been shown in Ref.~\cite{Giunti:2002xg} that also in this case
flavor neutrinos are described by states that are determined by the interaction process.
However, in this paper we avoid the complications of the wave packet description
by considering the plane-wave approximation in which
flavor neutrinos are described by superpositions of massive neutrino states
with definite energy and momentum.

In this paper we discuss the description of flavor neutrinos
produced or detected in weak interaction processes in which multiple neutrinos are involved.
We show that the different flavor neutrinos cannot be described by pure states,
but require a density matrix description
(note that there are other situations involving neutrinos which also require a density matrix description, such as when dealing with unpolarized beams; see, e.g.,
Refs.~\cite{Ochman:2007vn,Szafron:2011zz,Szafron:2012mi}).
We derive the appropriate density matrix and show that, under the appropriate approximations
the density matrix description leads to the standard oscillation probability
in the standard case of mixing of three light neutrinos.
We discuss also the effects of mixing of the three standard light neutrinos
with heavy neutrinos which can be either decoupled
because their masses are much larger than the
maximum neutrino energy in the neutrino production process
or because they are produced and detected incoherently
(see Refs.~\cite{Kayser:1981ye,Giunti:2003ax,Giunti:2007ry,Akhmedov:2009rb,Akhmedov:2017xxm}).

We consider also the more complicated case of neutrino-electron elastic scattering,
in which the flavors of the initial and final neutrinos are not determined,
but there is a flavor dependence,
because the
$\nu_{e}$ component interacts through both charged and neutral currents
whereas
the $\nu_{\mu}$ and $\nu_{\tau}$ components interact only through neutral currents.

This paper is organized as follows.
In Sec.~\ref{sec:II}
we briefly review the derivation of the flavor states
for processes involving only one flavor neutrino.
In Sec.~\ref{sec:IIA}
we discuss the effects of heavy neutrinos
in processes involving only one flavor neutrino.
In Sec.~\ref{sec:III} we present the density matrix description of the
$\nu_{e}$ and $\bar\nu_{\mu}$
produced in $\mu^+$ decay
as an example of a production process involving more than one neutrino.
In Sec.~\ref{sec:IV} we derive the density matrix description of flavor neutrinos
in detection processes considering the example of
$\nu_{\mu}$-electron scattering.
In Sec.~\ref{sec:IVb} we consider
neutrino-electron elastic scattering.
Finally, Sec.~\ref{sec:V} presents our conclusions.


\section{Flavor states for one-neutrino processes}\label{sec:II}

Flavor neutrinos states are commonly described as superpositions of neutrino states with a well-defined mass which amounts to a change of orthonormal basis using the unitary mixing matrix
that diagonalizes the mass matrix of the neutrino fields.
In quantum field theory, however, flavor neutrinos cannot be fundamentally described as excitations of flavor fields (due to the lack of a natural Fock space~\cite{Giunti:1991cb}) and are basically a phenomenological concept.
In this Section we give a short review of the derivation of the flavor states that describe flavor neutrinos produced or detected in weak interaction processes
\cite{Giunti:1991cb,Bilenky:2001yh,Giunti:2002xg,Giunti:2006fr,Giunti:2007ry}.

Let us consider the decay
\begin{eqnarray}
      P_I \to P_F + \bar{l}_\alpha + \nu_\alpha,
\label{decay}
\end{eqnarray}
where $P_I$ and $P_F$ stand for the initial and final particles besides the produced anti-lepton $\bar{l}_\alpha$
(with $\alpha \in \{e,\mu,\tau \}$) and its associated neutrino $\nu_\alpha$,
here understood as the superposition of states of massive neutrinos $\nu_i$
(with $i \in \{1,2,3 \}$)
which we derive explicitly below.

The final state of such decay is given by the action of the $\mathsf{S}$ matrix over the initial state $|i\rangle = | P_I \rangle$, i.e.,
\begin{eqnarray}
      |f \rangle &\propto&  \left( \mathsf{S}-\mathsf{I} \right)| i \rangle \nonumber \\
      &=& \sum_j \mathcal{A}^{\text{P}}_{\alpha,  j} |\nu_j,l^+_{\alpha},P_F \rangle + ...,
\end{eqnarray}
where we disregard the possibility of no decay and ``...'' denotes all other possible channels which do not concern us here. Since the other decays contained in ``...'' are orthogonal to
$|\nu_j,l^+_{\alpha},P_F \rangle$ and these states are orthonormal,
the coefficients $\mathcal{A}^{\text{P}}_{\alpha, j}$ (where $P$ stands for [neutrino] production)
are given by
\begin{eqnarray}
      \mathcal{A}^{\text{P}}_{\alpha, j} = \langle \nu_j, l^+_\alpha,P_F |\mathsf{S} | P_I \rangle. \label{amplitude}
\end{eqnarray}

The state describing the emitted flavor neutrino $\nu_{\alpha}$
is obtained by projecting the state $|f\rangle$ over $|l^+_\alpha, P_F\rangle$:
\begin{equation}
|\nu_\alpha \rangle
\propto
\langle l^+_\alpha, P_F | f \rangle
.
\label{projection}
\end{equation}
The resulting normalized flavor neutrino state is 
\begin{equation}
      |\nu_\alpha \rangle = \left(\sum_k |\mathcal{A}^{\text{P}}_{\alpha, k}|^2 \right)^{-1/2} \sum_j \mathcal{A}^{\text{P}}_{\alpha, j} |\nu_j\rangle. \label{weak-process-state}
\end{equation}
This state describes the neutrino produced in the decay (\ref{decay}) together with the
charged antilepton of the same flavor.
It is different from the standard flavor state,
\begin{equation}
    |\nu_\alpha \rangle_{\text{std}} = \sum_i U^*_{\alpha i} |\nu_i \rangle, \label{std_states}
\end{equation}
because the coefficients that determine
the superposition of the massive neutrinos
are the matrix elements (\ref{amplitude}) of the neutrino production process (\ref{decay}).
In order to show how they are related, we expand the $\mathsf{S}$ matrix up to first order
in the Fermi coupling constant $G_{\text{F}}$ as
\begin{eqnarray}
      \mathsf{S} &\approx& \mathsf{I} - i \, \frac{G_{\text{F}}}{\sqrt{2}} \int \text{d}^{4}x \,  \hat{j}_{\text{CC}\rho}^{\dagger}(x) \hat{j}_{\text{CC}}^\rho(x),
      \label{S_CC}
\end{eqnarray}
with the weak charged current $\hat{j}_{\text{CC}}^\rho(x)$ given by
\begin{eqnarray}
      \hat{j}_{\text{CC}}^\rho(x) = \sum_{\alpha,k} U^*_{\alpha k} \hat{\overline{\nu}}_{k}(x) \gamma^{\rho} \left( 1 - \gamma^{5} \right) \hat{l}_\alpha(x) + \hat{h}_{\text{CC}}^\rho(x),
\end{eqnarray}
where $\hat{h}_{\text{CC}}^\rho(x)$ is the hadronic part of the weak charged current.
Now, using Eq.~(\ref{amplitude}) we can write
\begin{eqnarray}
      \mathcal{A}^{\text{P}}_{\alpha, j} = U^*_{\alpha j} \mathcal{M}^{\text{P}}_{j},
\end{eqnarray}
where $\mathcal{M}^{\text{P}}_{j}$ is given by
\begin{align}
\mathcal{M}^{\text{P}}_{j} 
=
- i \, \frac{G_{\text{F}}}{\sqrt{2}} \int \text{d}^{4}x
\null & \null
\langle \nu_j l^+_\alpha |  \hat{\overline{\nu}}_{j}(x) \gamma^{\rho} \left( 1 - \gamma^{5} \right) \hat{l}_\alpha(x) | 0 \rangle
\nonumber
\\
\null & \null
\times J^{P_I \to P_F}_\rho(x),
\label{MP}
\end{align}
with the hadronic transition amplitude
\begin{equation}
    J^{P_I \to P_F}_\rho (x) \equiv \langle P_F | \hat{h}_\rho(x) | P_I \rangle.
\end{equation}
This decomposition allows us to expand Eq.~(\ref{weak-process-state}) as
\begin{equation}
|\nu_\alpha \rangle
=
\left(\sum_k |U_{\alpha k}|^2 |\mathcal{M}^{\text{P}}_{k}|^2 \right)^{-1/2}
\sum_j U^*_{\alpha j} \mathcal{M}^{\text{P}}_{j} |\nu_j\rangle
.
\label{wps}
\end{equation}

In the standard framework of three-neutrino mixing,
it is known that the neutrino masses are smaller than about 2 eV
(see Refs.~\cite{Giunti:2007ry,Bilenky:2018hbz,Patrignani:2016xqp}).
In this case,
the effects of the neutrino masses in the
matrix elements (\ref{MP}) are negligible
since in all neutrino oscillation experiments the neutrino energy is larger than about 0.1 MeV.
Hence,
in the standard framework of three-neutrino mixing
and in its extensions with extra neutrinos much lighter than 0.1 MeV,
we can approximate
$ \mathcal{M}^{\text{P}}_{j} \simeq \mathcal{M}^{\text{P}}_{0} $ $\forall j$,
where
$\mathcal{M}^{\text{P}}_{0}$
is the value of the matrix element in the Standard Model with massless neutrinos.
Then,
using the unitarity of the mixing matrix,
we obtain the standard states given in Eq.~(\ref{std_states}).
Assuming a similar description of the detected flavor neutrinos
in a neutrino oscillation experiment,
one then obtains the standard flavor transition probability
\begin{equation}
\mathsf{P}_{\nu_{\alpha} \to \nu_{\beta}}(L,E)
=
\sum_{j,k}  U^*_{{\alpha}j} U_{{\beta}j} U_{{\alpha}k} U^*_{{\beta}k}
\exp{\left( -i\frac{\Delta m^2_{jk} L}{2E} \right)}
,
\label{Pstd}
\end{equation}
where $\Delta m^2_{jk} = m^2_j-m^2_k$, $E$ is the neutrino energy
and
$L$
is the source-detector distance.

The flavor neutrino states allow us to give a complete picture of neutrino production, oscillations and detection in the case of processes involving a single neutrino \cite{Giunti:2002xg}.
Particularly notable is that they give the correct production rate at the neutrino source as an incoherent sum of massive neutrino states
\cite{Shrock:1980vy,McKellar:1980cn,Kobzarev:1980nk,Shrock:1980ct,Shrock:1981wq}.


\section{Effects of heavy neutrinos}\label{sec:IIA}

In the standard case of mixing of three light neutrinos
the formalism reviewed in Section~\ref{sec:II} is mainly interesting for the purely theoretical
purpose of deriving the flavor neutrino states
and the oscillation probability
from first principles.
On the other hand,
the flavor neutrino states (\ref{wps})
are practically useful in the presence of heavy neutrinos,
for which the mass effects in the corresponding
matrix elements (\ref{MP}) are not negligible.
A particularly simple and realistic case is that of mixing
between light neutrinos and
very heavy neutrinos
which are decoupled because their masses are much larger than the
maximum neutrino energy in the neutrino production process.
This situation occurs, for example, in see-saw models
\cite{Minkowski:1977sc,Yanagida:1979as,GellMann:1980vs,Mohapatra:1980ia}.

Let us consider the case of
$N_{\ell}$ light and $N_{d}$ heavy decoupled neutrinos.
Ordering the index of the massive neutrinos according to the size of their masses,
we have
$\mathcal{M}^{\text{P}}_{j} = 0$ for $j>N_{\ell}$.
Considering
$\mathcal{M}^{\text{P}}_{j} \simeq \mathcal{M}^{\text{P}}_{0}$ for $j \leq N_{\ell}$,
we obtain directly from Eq.~(\ref{weak-process-state}) the flavor neutrino states
\begin{equation}
|\nu_\alpha \rangle
=
\left(\sum_{k \leq N_{\ell}} |U_{\alpha k}|^2\right)^{-1/2}
\sum_{j \leq N_{\ell}} U^*_{\alpha j} |\nu_j\rangle
.
\label{heavy1}
\end{equation}
These flavor neutrino states can be written in the same form
of the standard flavor states in Eq~(\ref{std_states})
by defining the coefficients
$
\tilde{U}_{\alpha j}
\equiv
U_{\alpha j} / \sqrt{ \sum_{k \leq N_{\ell}} |U_{\alpha k}|^2 }
$
(see, for example, Ref.~\cite{Antusch:2006vwa}).
However,
the new coefficients
$ \tilde{U}_{\alpha j} $
are different from the elements of the mixing matrix of the neutrino fields
and do not constitute a unitary matrix.

Notice that,
albeit the flavor neutrino states that we have derived are normalized by construction
($ \langle \nu_\alpha | \nu_\alpha \rangle = 1 $),
they may be non-orthogonal.
Indeed in the presence of decoupled heavy neutrinos,
from Eq.~(\ref{heavy1}),
for $\alpha\neq\beta$ we have
\begin{equation}
\langle \nu_\beta | \nu_\alpha \rangle
=
\frac{ \sum_{j \leq N_{\ell}} U_{\beta j} U^*_{\alpha j} }{
\sqrt{ \left( \sum_{k \leq N_{\ell}} |U_{\beta k}|^2 \right)
       \left( \sum_{k' \leq N_{\ell}} |U_{\alpha k'}|^2 \right) }
}
.
\label{heavy2}
\end{equation}
This means that there is a ``zero-distance''
probability to detect the neutrino with a flavor which is different from the production flavor\footnote{
In other words, the produced neutrino is a superposition of
massive neutrinos that in a charged-current weak interaction can generate
different charged leptons.
}
\cite{Langacker:1988up}.
This effect is not inconsistent with the
unitary normalization of the states ($ \langle \nu_\alpha | \nu_\alpha \rangle = 1 $)
which quantifies the obvious fact that a neutrino which is produced with a flavor $\alpha$
can be detected with the same flavor $\alpha$ immediately after production
without any suppression due to neutrino mixing.
It is apparently puzzling that
$ \sum_{\beta} | \langle \nu_\beta | \nu_\alpha \rangle |^2 \geq 1 $,
which seems a violation of unitarity.
However,
one must take into account that the experimental event rate
is given by
(the constant of proportionality depends on the size and composition
of the source and detector)
\begin{equation}
R_{\alpha\beta}(L,E)
\propto
\Gamma_{\alpha}(E)
\mathsf{P}_{\nu_{\alpha}\to\nu_{\beta}}(L,E)
\sigma_{\beta}(E)
,
\label{rate1}
\end{equation}
where
$\Gamma_{\alpha}(E)$
and
$\sigma_{\beta}(E)$
are, respectively,
the rate of $\nu_{\alpha}$ production in the source
and
the detection cross-section of $\nu_{\beta}$,
which must be calculated taking into account the effects of the heavy neutrino masses.
In the case of heavy decoupled neutrinos we have
\begin{align}
\null & \null
\Gamma_{\alpha}(E)
=
\Gamma_{\alpha}^{0}(E)
\left( \sum_{k \leq N_{\ell}} |U_{\alpha k}|^2 \right)
,
\label{Gamma}
\\
\null & \null
\sigma_{\beta}(E)
=
\sigma_{\beta}^{0}(E)
\left( \sum_{k \leq N_{\ell}} |U_{\beta k}|^2 \right)
,
\label{sigma}
\end{align}
where
$\Gamma_{\alpha}^{0}(E)$
and
$\sigma_{\beta}^{0}(E)$
are the values of the production rate and detection cross section
for massless neutrinos that are normally used in the calculations
of the rates in neutrino oscillation experiments.
From Eq.~(\ref{heavy1}) we obtain the flavor transition probability
\begin{align}
\mathsf{P}_{\nu_{\alpha} \to \nu_{\beta}}
\null & \null
(L,E)
=
\left(\sum_{j' \leq N_{\ell}} |U_{\alpha j'}|^2\right)^{-1}
\left(\sum_{k' \leq N_{\ell}} |U_{\beta k'}|^2\right)^{-1}
\nonumber
\\
\null & \null
\times
\sum_{j,k\leq N_{\ell}}  U^*_{{\alpha}j} U_{{\beta}j} U_{{\alpha}k} U^*_{{\beta}k}
\exp{\left( -i\frac{\Delta m^2_{jk} L}{2E} \right)}
.
\label{heavyprob1}
\end{align}
Therefore,
the coefficients
in Eqs.~(\ref{Gamma}), (\ref{sigma}), and (\ref{heavyprob1})
cancel and
the experimental event rate can be written conveniently in the usual form
\begin{equation}
R_{\alpha\beta}(L,E)
\propto
\Gamma_{\alpha}^{0}(E)
\mathsf{P}_{\nu_{\alpha}\to\nu_{\beta}}^{\text{eff}}(L,E)
\sigma_{\beta}^{0}(E)
,
\label{rate2}
\end{equation}
with the effective flavor transition probability
\begin{equation}
\mathsf{P}_{\nu_{\alpha}\to\nu_{\beta}}^{\text{eff}}(L,E)
=
\sum_{j,k\leq N_{\ell}} U^*_{{\alpha}j} U_{{\beta}j} U_{{\alpha}k} U^*_{{\beta}k}
\exp{\left( -i\frac{\Delta m^2_{jk} L}{2E} \right)}
.
\label{heavyprob2}
\end{equation}
In this way,
one can easily take into account the mixing with decoupled heavy neutrinos
in the usual method of calculation of the
rates in neutrino oscillation experiments
by an appropriate choice of the normalization of the flavor transition probability.

Moreover, expression (\ref{rate2}) for the experimental rate
can be easily generalized to the case of heavy neutrinos which are not decoupled,
but are produced and detected incoherently
(see Refs.~\cite{Kayser:1981ye,Giunti:2003ax,Giunti:2007ry,Akhmedov:2009rb,Akhmedov:2017xxm}).
Here and in the following we assume, for simplicity, that these heavy neutrinos
have mass splittings of the order or larger than their masses,
in order to be produced incoherently among themselves.
This is a likely situation,
since having almost degenerate heavy neutrinos would require an appropriate fine tuning.

Heavy neutrinos are produced incoherently
if the mass differences are larger than the mass uncertainty
in the production process.
From the relativistic energy momentum dispersion relation,
one can find that for ultrarelativistic neutrinos the mass uncertainty
in the production process is of the order of
$\sqrt{E \sigma_{E}}$,
where $E \simeq p$ is the neutrino energy approximately equal to its momentum and
$\sigma_{E} \simeq \sigma_{p}$
is the energy-momentum uncertainty in the production process
(see Refs.~\cite{Kayser:1981ye,Giunti:2003ax,Giunti:2007ry,Akhmedov:2009rb,Akhmedov:2017xxm}).
For neutrinos produced by decays in matter the energy uncertainty is determined by the
spatial localization
$\sigma_{x}$
of the production process
through the uncertainty principle:
$ \sigma_{E} \simeq \sigma_{p} \sim \sigma_{x}^{-1} $.
In normal matter $\sigma_{x}$ is of the order of the interatomic distance,
$\sigma_{x} \sim 10^{-8} \, \text{cm}$,
leading to
$ \sigma_{E} \sim 1 \, \text{keV} $.
Considering a typical energy of 10 MeV,
neutrinos with masses larger than 100 keV are produced incoherently.
For the detection process one can make similar considerations,
but it is often more important that coherence is effectively lost because of the separation of the wave packets, which occurs quickly in the case of heavy neutrinos.
Indeed, since the size of the neutrino wave packets is of the order of $\sigma_{x}$,
the coherence length is given by
$L_{\text{coh}} \sim E^2 \sigma_{x} / | \Delta{m}^2 |$
(see Refs.~\cite{Giunti:2003ax,Giunti:2007ry,Akhmedov:2009rb,Akhmedov:2017xxm}),
and for neutrinos produced in matter with $E \sim 10 \, \text{MeV}$
the coherence length is smaller than about 1 m if their masses are larger than about 100 eV.
Since all neutrino oscillation experiments have a source-detector distance larger than 1 m,
if the neutrino energy is smaller than about 10 MeV
the heavy neutrinos with masses larger than about 100 eV must be treated incoherently.

Let us consider
$N_{\ell}$ light neutrinos,
$N_{h}$ heavy incoherent non-decoupled neutrinos,
and $N_{d}$ heavy decoupled neutrinos.
Ordering the index of the massive neutrinos according to the size of their masses,
we have
\begin{align}
R_{\alpha\beta}(L,E)
\propto
\null & \null
\Gamma_{\alpha}^{0}(E)
\mathsf{P}_{\nu_{\alpha}\to\nu_{\beta}}^{\text{eff}}(L,E)
\sigma_{\beta}^{0}(E)
\nonumber
\\
\null & \null
+
\sum_{k=N_{\ell}+1}^{N_{\ell}+N_{h}}
\Gamma_{\alpha}^{k}(E)
|U_{\alpha k}|^2
|U_{\beta k}|^2
\sigma_{\beta}^{k}(E)
,
\label{rate3}
\end{align}
where
$\Gamma_{\alpha}^{k}(E)$
and
$\sigma_{\beta}^{k}(E)$
are the production rate and the detection cross section for a neutrino with heavy mass
$m_{k}$
which is produced and detected incoherently.

Finally, let us also notice that the effective flavor transition probability (\ref{heavyprob2})
is more attractive than the flavor transition probability (\ref{heavyprob1})
because
\begin{equation}
\sum_{\beta}
\mathsf{P}_{\nu_{\alpha}\to\nu_{\beta}}^{\text{eff}}(L,E)
=
\sum_{j\leq N_{\ell}} |U_{{\alpha}j}|^2
\leq
1
,
\label{heavyprob2sum}
\end{equation}
whereas
\begin{equation}
\sum_{\beta}
\mathsf{P}_{\nu_{\alpha}\to\nu_{\beta}}(L,E)
=
\left( \sum_{j\leq N_{\ell}} |U_{{\alpha}j}|^2 \right)^{-1}
\geq
1
,
\label{heavyprob1sum}
\end{equation}
which follows from the above-mentioned fact that
$ \sum_{\beta} | \langle \nu_\beta | \nu_\alpha \rangle |^2 \geq 1 $.
Note that in Eqs.~(\ref{heavyprob2sum}) and (\ref{heavyprob1sum})
the summation over the flavor index $\beta$
is carried out over all the neutrino favors, including the three active neutrinos
and the sterile neutrinos that are present
if the total number of massive neutrinos is bigger than three.


\section{Multiple neutrino production}\label{sec:III}

The construction of the flavor neutrino states reviewed in Section~\ref{sec:II}
is well-suited, for instance, for the description of
$\nu_{e}$'s produced in $\beta^{+}$ decays
or
$\nu_{\mu}$'s produced in $\pi^{+}$ decays
(and,
with obvious modifications,
$\bar{\nu}_{e}$'s produced in $\beta^{-}$ decays
or
$\bar\nu_{\mu}$'s produced in $\pi^{-}$ decays).
In these cases we have only one flavor neutrino which is described by a pure state.

Let us now discuss how the description of flavor neutrinos
must be modified in the case of reactions involving more than one neutrino.

Let us consider as an example the neutrino creation process
\begin{equation}
    \mu^{+} \to e^{+} + \nu_{e} + \bar\nu_{\mu}. \label{mupdecay}
\end{equation}
How to describe the emitted electron neutrino and muon antineutrino?
One can see that the method reviewed in Sec.~\ref{sec:I} must be extended
by considering, for example, the description of the $\nu_{e}$.
The derivation of its state would require the projection
$\langle e^+, \bar\nu_{\mu} | f \rangle$
analogous to Eq.~(\ref{projection}),
which in this case is not possible, because the state $|\bar\nu_{\mu}\rangle$
is still not defined.

In order to derive the correct description of the emitted
$\nu_{e}$ and $\bar\nu_{\mu}$
let us first consider the final state of the process (\ref{mupdecay}):
\begin{equation}
    |f\rangle \propto \left(\mathsf{S}-\mathsf{I} \right) \, | \mu^{+} \rangle = \sum_{k,j} \mathcal{A}^{\text{P}}_{\mu,e;k,j} \, | e^{+}, \nu_{k}, \bar\nu_{j} \rangle + ...,  \label{finalstate}
\end{equation}
where ``...'' denotes other possible decay channels irrelevant for our purposes
(e.g., $\mu^+ \to e^+ + \gamma$).
Using the orthonormality of the states,
the coefficients $\mathcal{A}^{\text{P}}_{\mu,e;k,j}$ are given by
\begin{equation}
    \mathcal{A}^{\text{P}}_{\mu,e;k,j} = \langle e^{+}, \nu_{k}, \bar\nu_{j} | \mathsf{S} | \mu^{+} \rangle = U^{*}_{e k} U_{\mu j} \mathcal{M}^{\text{P}}_{k,j}, \label{AP}
\end{equation}
where
\begin{align}
\mathcal{M}^{\text{P}}_{k,j}
=
-i \, \frac{G_{\text{F}}}{\sqrt{2}}
\int 
\null & \null
\text{d}^{4}x
\langle e^{+}, \nu_{k}, \bar{\nu}_{j} | \hat{\overline{\nu}}_{k}(x) \gamma^{\rho}
\left( 1 - \gamma^{5} \right) \hat{e}(x)
\nonumber
\\
\null & \null
\times
\hat{\overline{\mu}}(x) \gamma_{\rho} \left( 1 - \gamma^{5} \right) \hat{\nu}_{j}(x) | \mu^{+}  \rangle. \label{Mkj}
\end{align}
The final state in Eq.~(\ref{finalstate}) is an entangled state in which it is not possible to separate the neutrino and antineutrino components. Therefore, the neutrino and the antineutrino cannot be described by a pure state and should each separately be described by a density matrix, which is the most general description of a quantum system which may be a subsystem of a larger closed system.

The density matrix operator that describes the complete final state in Eq.~(\ref{finalstate}) is
\begin{align}
\hat{\rho} 
=
\null & \null
| f \rangle \langle f |
\nonumber
\\
=
\null & \null
N^{\text{P}}
\sum_{k,j,k',j'}
\mathcal{A}^{\text{P}}_{\mu,e;k,j} \mathcal{A}_{\mu,e;k',j'}^{\text{P}*}
| e^{+}, \nu_{k}, \bar\nu_{j} \rangle \langle e^{+}, \nu_{k'}, \bar\nu_{j'} |,
\label{desmat}
\end{align}
where
$N^{\text{P}}$ is a normalization coefficient determined by the condition $\operatorname{Tr}(\hat{\rho})=1$, i.e.,
\begin{equation}
    \sum_{k'',j''} \langle e^{+}, \nu_{k''}, \bar\nu_{j''} | \hat{\rho} | e^{+}, \nu_{k''}, \bar\nu_{j''} \rangle = 1. \label{normcond}
\end{equation}
This gives
\begin{eqnarray}
    N^{\text{P}} &=& \left( \sum_{k,j} | \mathcal{A}^{\text{P}}_{\mu,e;k,j} |^2 \right)^{-1} \nonumber \\
    &=& \left(\sum_{k,j}  |U_{e k}|^2 |U_{\mu j}|^2 |\mathcal{M}^{\text{P}}_{k,j}|^2 \right)^{-1}. \label{norm}
\end{eqnarray}
The $\nu_{e}$ and $\bar\nu_{\mu}$ are separately described by the partial traces over the other degrees of freedom of the complete system:
\begin{eqnarray}
    \hat{\rho}_{\nu_{e}} &=& \sum_{j} \langle e^{+}, \bar\nu_{j} | \hat{\rho} | e^{+}, \bar\nu_{j} \rangle \nonumber \\
    &=& N^{\text{P}} \sum_{k,k',j} \mathcal{A}^{\text{P}}_{\mu,e;k,j} \mathcal{A}_{\mu,e;k',j}^{\text{P}*} \, | \nu_{k} \rangle \langle \nu_{k'} |, \label{dmnue1} \\
    \hat{\rho}_{\bar\nu_{\mu}} &=& \sum_{k} \langle e^{+}, \nu_{k} | \hat{\rho} | e^{+}, \nu_{k} \rangle \nonumber \\
    &=& N^{\text{P}} \sum_{k,j,j'} \mathcal{A}^{\text{P}}_{\mu,e;k,j} \mathcal{A}_{\mu,e;k,j'}^{\text{P}*} \, | \bar\nu_{j} \rangle \langle \bar\nu_{j'} |. \label{dmnum1}
\end{eqnarray}
Using Eq.~(\ref{AP}), we obtain
\begin{align}
    \hat{\rho}_{\nu_{e}} = \null & \null N^{\text{P}} \sum_{j} |U_{\mu j}|^2 \sum_{k,k'} U_{e k}^{*} U_{e k'} \mathcal{M}^{\text{P}}_{k,j} \mathcal{M}_{k',j}^{\text{P}*} \, | \nu_{k} \rangle \langle \nu_{k'} |, \label{dmnue2} \\
    \hat{\rho}_{\bar\nu_{\mu}} = \null & \null N^{\text{P}} \sum_{k} |U_{e k}|^2 \sum_{j,j'} U_{\mu j} U_{\mu j'}^{*} \mathcal{M}^{\text{P}}_{k,j} \mathcal{M}_{k,j'}^{\text{P}*} \, | \bar\nu_{j} \rangle \langle \bar\nu_{j'} |. \label{dmnum2}
\end{align}
These density matrices describe separately the $\nu_{e}$ and $\bar\nu_{\mu}$
and one can see that,
because of the dependence of the interaction matrix elements
on $j$ in Eq.~(\ref{dmnue2})
and
on $k$ in Eq.~(\ref{dmnum2}),
they are not density matrices of pure states
(as one can also verify by checking that
$\text{Tr}(\hat{\rho}_{\nu_{e}}^2) < 1$
and
$\text{Tr}(\hat{\rho}_{\bar\nu_{\mu}}^2) < 1$).
Hence,
the $\nu_{e}$ and $\bar\nu_{\mu}$
cannot be described by pure states. It is interesting to note that while
the complete density matrix in Eq.~(\ref{desmat}) allows us to correctly calculate the decay rate for the process (\ref{mupdecay}) as an incoherent sum over massive neutrino states,
the density matrices in Eqs.~(\ref{dmnue2}) and (\ref{dmnum2}) do not.
This can be understood as a consequence of information loss due to taking the partial trace of the complete density matrix.

The $\nu_{e}$ and $\bar\nu_{\mu}$ can be approximately described by pure states
in experiments which are not sensitive to the dependence of $\mathcal{M}^{\text{P}}_{k,j}$ on the neutrino masses, where it is possible to approximate $\mathcal{M}^{\text{P}}_{k,j} \simeq \mathcal{M}^{\text{P}}$, $\forall j,k$. Taking into account that in this case $(N^{\text{P}})^{-1} \simeq |\mathcal{M}^{\text{P}}|^2$, we obtain
\begin{align}
    \hat{\rho}_{\nu_{e}} \simeq \null & \null \left( \sum_{k} U_{e k}^{*} | \nu_{k} \rangle \right) \left( \sum_{k'} U_{e k'} \langle \nu_{k'} | \right), \label{dmnue3} \\
    \hat{\rho}_{\bar\nu_{\mu}} \simeq \null & \null \left( \sum_{j} U_{\mu j} | \bar\nu_{j} \rangle \right) \left( \sum_{j'} U_{\mu j'}^{*} \langle \bar\nu_{j'} | \right), \label{dmnum3}
\end{align}
which are the density matrices associated with the standard flavor states given in Eq.~(\ref{std_states}).
Therefore, in this approximation we recover the standard description
of the neutrino flavor states.

Now consider, for example, the electron neutrino, described by $\hat{\rho}_{\nu_e}$, as the initial state of an oscillation experiment.
It propagates freely and then it can be detected with a different flavor,
for instance via the process
\begin{equation}
    \nu_\mu + D_I \to \mu^- + D_F \label{detection},
\end{equation}
where $D_I$ and $D_F$ are the initial and final states of the other particles involved in the detection.
We first apply a spatio-temporal translation, $\mathcal{U}(t=T,\vec{x}=\vec{L}) = \exp{\left(-i \hat{p}^0 T + i \vec{\hat{p}} \cdot \vec{L} \right)}$ to the density matrix $\hat{\rho}_{\nu_e}$,
which gives
\begin{eqnarray}
    \hat{\rho}_{\nu_e}(T,\vec{L}) &=& \mathcal{U}(T,\vec{L}) \hat{\rho}_{\nu_e} \mathcal{U}^\dagger(T,\vec{L}) \nonumber \\
    &=& N^{\text{P}} \sum_j |U_{\mu j}|^2 \sum_{k,k'} U^*_{e k} U_{e k'} \mathcal{M}^{\text{P}}_{k,j} \mathcal{M}^{\text{P}*}_{k',j} \nonumber \\
    && \times \exp \left[ -i(E_k-E_{k'})T \vphantom{\vec{L}}  \right. \nonumber \\ 
    &&\left. + \ i (\vec{p}_k-\vec{p}_{k'}) \cdot \vec{L} \right] |\nu_k \rangle \langle \nu_{k'}|.  
\end{eqnarray}
In the relativistic approximation, where $T \simeq L \equiv |\vec{L}|$,
considering for simplicity all the massive neutrinos propagating in the direction of $\vec{L}$
(see the discussion in Section~8.1.3 of Ref.~\cite{Giunti:2007ry})
and
using Eq.~(5.7) of Ref.~\cite{Giunti:2006fr},
we obtain
\begin{eqnarray}
    \hat{\rho}_{\nu_e}(T,\vec{L}) &=& \mathcal{U}(T,\vec{L}) \hat{\rho}_{\nu_e} \mathcal{U}^\dagger(T,\vec{L}) \nonumber \\
    &=& N^{\text{P}} \sum_j |U_{\mu j}|^2 \sum_{k,k'} U^*_{e k}U_{e k'} \mathcal{M}^{\text{P}}_{k,j}\mathcal{M}^{P*}_{k',j} \nonumber \\
    && \times \exp{\left(-i \frac{\Delta m^2_{k k'}L}{2E} \right)} |\nu_k \rangle \langle \nu_k'|,
\label{rho_nue}
\end{eqnarray}
where $\Delta m^2_{k k'} = m^2_k - m^2_{k'}$
and
$E$ is the neutrino energy in the massless approximation.
Let us now consider the probability of detecting the neutrino as a $\nu_{\mu}$,
\begin{equation}
    \mathsf{P}_{\nu_e \to \nu_\mu} =  \text{Tr} \left[ \hat{\rho}_{\nu_e}(T,\vec{L}) |\nu^\text{D}_\mu \rangle \langle \nu^\text{D}_\mu | \right],
\label{probem}
\end{equation}
where the state $|\nu^\text{D}_\mu \rangle$ is given by Eq.~(\ref{weak-process-state}) with the appropriate amplitudes associated with the detection (D) reaction in Eq.~(\ref{detection}).  
The resulting probability is 
\begin{eqnarray}
    \mathsf{P}_{\nu_e \to \nu_\mu}&=& \sum_{j,k,k'} |U_{\mu j}|^2 \left[\frac{ \mathcal{M}^\text{D}_{k} \mathcal{M}^{\text{D}*}_{k'}}{\left(\sum_i  |U_{\mu i}|^2 |\mathcal{M}^\text{D}_{i}|^2 \right)} \right] \nonumber \\
    && \times \left[\frac{ \mathcal{M}^{\text{P}}_{k, j} \mathcal{M}^{\text{P}*}_{k',j}}{\left(\sum_{a,b}  |U_{e a}|^2 |U_{\mu b}|^2 |\mathcal{M}^{\text{P}}_{a, b}|^2 \right)} \right] \nonumber \\
    && \times U^*_{e k} U_{\mu k} U_{e k'} U^*_{\mu k'} \exp{\left( -i\frac{\Delta m^2_{k k'}L}{2E} \right)}, \ \ \ \ \ 
\end{eqnarray}
which has the standard oscillation phase.
Moreover,
if the differences of the neutrino masses are negligible in the production and detection processes,
we have
$\mathcal{M}^{\text{D}}_{k} \simeq \mathcal{M}^{\text{D}}$
and
$\mathcal{M}^{\text{P}}_{k,j} \simeq \mathcal{M}^{\text{P}}$,
$\forall j,k$,
which leads to the standard flavor transition probability (\ref{Pstd})
with $\alpha=e$ and $\beta=\mu$.

It is also interesting to consider the case of
$N_{\ell}$ light and $N_{d}$ heavy decoupled neutrinos
discussed in Section~\ref{sec:IIA}.
Since in this case
$\mathcal{M}^{\text{P}}_{k,j} = 0$
for $k>N_{\ell}$ and/or $j>N_{\ell}$
and
we can approximate
$\mathcal{M}^{\text{P}}_{k,j} \simeq \mathcal{M}^{\text{P}}_{0,0}$
for
$k,j \leq N_{\ell}$,
from Eqs.~(\ref{dmnue2}) and (\ref{dmnum2}) we obtain
\begin{align}
\hat{\rho}_{\nu_{e}}
\simeq
\null & \null
\left( \sum_{j \leq N_{\ell}} |U_{e j}|^2 \right)^{-1}
\left( \sum_{k \leq N_{\ell}} U_{e k}^{*} | \nu_{k} \rangle \right)
\nonumber
\\
\null & \null
\hspace{2.4cm}
\times
\left( \sum_{k' \leq N_{\ell}} U_{e k'} \langle \nu_{k'} | \right),
\label{dmnue4}
\\
\hat{\rho}_{\bar\nu_{\mu}}
\simeq
\null & \null
\left( \sum_{j \leq N_{\ell}} |U_{\mu k}|^2 \right)^{-1}
\left( \sum_{j \leq N_{\ell}} U_{\mu j} | \bar\nu_{j} \rangle \right)
\nonumber
\\
\null & \null
\hspace{2.4cm}
\times
\left( \sum_{j' \leq N_{\ell}} U_{\mu j'}^{*} \langle \bar\nu_{j'} | \right), \label{dmnum4}
\end{align}
which are the density matrices associated with the flavor states in Eq.~(\ref{heavy1}).
Hence, also in this case we recover the description in terms of
flavor states.
Note however, that the result is not trivial,
in particular regarding the disappearance of any effect due to the mixing of the undetected neutrino.

On the other hand,
in the case of
$N_{\ell}$ light neutrinos,
$N_{h}$ heavy neutrinos which are produced and detected incoherently,
and $N_{d}$ heavy decoupled neutrinos,
from Eq.~(\ref{dmnue2}) applied to this situation,
we obtain the rather complicated density matrix
\begin{align}
\hat{\rho}_{\nu_{e}}
=
\null & \null
N^{\text{P}}
\Bigg[
\Bigg(
|\mathcal{M}^{\text{P}}_{0,0}|^2 \sum_{j \leq N_{\ell}} |U_{\mu j}|^2
\nonumber
\\
\null & \null
\hspace{-0.3cm}
+
\sum_{j=N_{\ell}+1}^{N_{\ell}+N_{h}} |U_{\mu j}|^2 |\mathcal{M}^{\text{P}}_{0,j}|^2
\Bigg)
\sum_{k,k' \leq N_{\ell}} U_{e k}^{*} U_{e k'} \, | \nu_{k} \rangle \langle \nu_{k'} |
\nonumber
\\
\null & \null
\hspace{-0.3cm}
+
\sum_{j \leq N_{\ell}+N_{h}} |U_{\mu j}|^2
\sum_{k=N_{\ell}+1}^{N_{\ell}+N_{h}} |U_{e k}|^2 |\mathcal{M}^{\text{P}}_{k,j}|^2 \, | \nu_{k} \rangle \langle \nu_{k} |
\Bigg]
.
\label{inc1}
\end{align}
Taking into account that the $\mu^{+}$ decay probability is $(N^{\text{P}})^{-1}$,
given in Eq.~(\ref{norm}),
and the probability of the detection process (\ref{detection}) is
$
(N^{\text{D}})^{-1}
=
\sum_{k} |U_{\mu k}|^2 |\mathcal{M}^\text{D}_{k}|^2
$,
the rate of a $\nu_{e}\to\nu_{\mu}$ oscillation experiment
is given by
\begin{equation}
R_{e\mu}(L,E)
\propto
\int \text{dPS}
\,
(N^{\text{P}})^{-1}
\,
\mathsf{P}_{\nu_e \to \nu_\mu}
\,
(N^{\text{D}})^{-1}
,
\label{ratem}
\end{equation}
where $\mathsf{P}_{\nu_e \to \nu_\mu}$ is the oscillation probability given by Eq.~(\ref{probem})
and
the integration over $\text{dPS}$ represents schematically the integration over the phase space.
After simplification of the factor
$N^{\text{P}}$ in $\hat{\rho}_{\nu_{e}}$
and the factor
$N^{\text{D}}$ in $|\nu^\text{D}_\mu \rangle \langle \nu^\text{D}_\mu |$,
integrating over the phase space
we obtain
\begin{align}
R_{e\mu}(L,E)
\propto
\null & \null
\left( \sum_{j \leq N_{\ell}} |U_{\mu j}|^2 \right)
\Gamma_{\mu^{+}}^{0,0}(E)
\mathsf{P}_{\nu_{e}\to\nu_{\mu}}^{\text{eff}}(L,E)
\sigma_{\mu}^{0}(E)
\nonumber
\\
\null & \null
\hspace{-1.5cm}
+
\sum_{j=N_{\ell}+1}^{N_{\ell}+N_{h}} |U_{\mu j}|^2
\Gamma_{\mu^{+}}^{0,j}(E)
\mathsf{P}_{\nu_{e}\to\nu_{\mu}}^{\text{eff}}(L,E)
\sigma_{\mu}^{0}(E)
\nonumber
\\
\null & \null
\hspace{-1.5cm}
+
\sum_{j \leq N_{\ell}+N_{h}} |U_{\mu j}|^2
\sum_{k=N_{\ell}+1}^{N_{\ell}+N_{h}}
\Gamma_{\mu^{+}}^{k,j}(E)
|U_{e k}|^2
|U_{\mu k}|^2
\sigma_{\mu}^{k}(E)
,
\label{inc2}
\end{align}
where
$\mathsf{P}_{\nu_{e}\to\nu_{\mu}}^{\text{eff}}(L,E)$
is the effective probability in Eq.~(\ref{heavyprob2}),
$\Gamma_{\mu^{+}}^{k,j}(E)$
is the decay rate of $\mu^{+} \to e^{+} + \nu_{k} + \bar\nu_{j}$,
and
$\sigma_{\mu}^{k}(E)$
is the detection cross section for a neutrino with mass
$m_{k}$.
For these quantities, $k=0$ and $j=0$ indicate massless neutrinos.
Equation~(\ref{inc2})
shows that the experimental rate depends not only on the heavy neutrino masses
which constitute the detected neutrino,
but also on those which constitute the undetected neutrino.


\section{Detection processes}\label{sec:IV}

It is interesting to study detection processes for flavor neutrinos where there is more than one neutrino involved
with the approach described above for production processes.
There are subtle differences which we discuss in this Section.

Let us consider as an example the ``inverse muon decay'' neutrino detection process
\begin{equation}
    \nu_{\mu} + e^{-} \to \mu^{-} + \nu_{e}. \label{imd}
\end{equation}
Although this process can be used to detect muon neutrinos \cite{Vilain:1996yf,Mishra:1990yf,Formaggio:2001jz}  it is not used in practice for neutrino oscillation experiments, because the neutrino energy threshold is high (about $10.92$ GeV) and the cross section is about one thousand times smaller than that of $\nu_{\mu}$ charged-current scattering on neutrons. However, at least in principle one can ask which is the correct description of the detected $\nu_{\mu}$, taking into account that the $\nu_{e}$ in the final state is a superposition of massive neutrinos which is not known {\it a priori}.

Since the final neutrino is a superposition of orthogonal massive neutrinos, the cross section of the process (\ref{imd}) is the incoherent sum of the cross sections with the different massive neutrinos in the final state:
\begin{equation}
    \sigma( \nu_{\mu} + e^{-} \to \mu^{-} + \nu_{e} ) = \sum_{j} \sigma( \nu_{\mu} + e^{-} \to \mu^{-} + \nu_{j} ). \label{csimd}
\end{equation}
Therefore, the detected $\nu_{\mu}$ must be described by a density matrix, which allows us to describe the incoherent sum in Eq.~(\ref{csimd}). We start by considering the separate processes
\begin{equation}
    \nu_{\mu} + e^{-} \to \mu^{-} + \nu_{j}. \label{imdj}
\end{equation}
The corresponding initial states are given by
\begin{equation}
    |i_{j}\rangle \propto \left(\mathsf{S}^{\dagger}-\mathsf{I} \right) | \mu^{-}, \nu_{j} \rangle = \sum_{k} \mathcal{A}^{\text{D}}_{\mu,e;k,j} \, | \nu_{k}, e^{-} \rangle + \ldots, \label{inij}
\end{equation}
with
\begin{equation}
    \mathcal{A}^{\text{D}}_{\mu,e;k,j} = \langle \nu_{k}, e^{-} | \mathsf{S}^{\dagger} | \mu^{-}, \nu_{j} \rangle = U_{\mu k}^{*} U_{ej} \mathcal{M}^{\text{D}}_{k,j}, \label{ADkj}
\end{equation}
where
\begin{eqnarray}
    \mathcal{M}^{\text{D}}_{k,j} &=& i \, \frac{G_{\text{F}}}{\sqrt{2}} \int \text{d}^{4}x \, \langle \nu_{k}, e^{-} | \overline{e}(x) \gamma^{\rho} \left( 1 - \gamma^{5} \right) \nu_{j}(x) \nonumber \\
    && \times \overline{\nu_{k}}(x) \gamma_{\rho} \left( 1 - \gamma^{5} \right) \mu(x) \, | \mu^{-}, \nu_{j} \rangle. \label{MDkj}
\end{eqnarray}

The density matrix operator that describes the initial state in the process (\ref{imd}) is then
\begin{eqnarray}
    \hat{\rho}^{\text{D}} &=& \frac{1}{3} \sum_{j} | i_{j} \rangle \langle i_{j} | \nonumber \\
    &=& N^{\text{D}}  \sum_{j,k,k'} \mathcal{A}^{\text{D}}_{\mu,e;k,j} \mathcal{A}^{\text{D}*}_{\mu,e;k',j} \, | \nu_{k}, e^{-} \rangle \langle \nu_{k'}, e^{-} |, \label{dmD}
\end{eqnarray}
where $N^\text{D}$ is the normalization coefficient given by
\begin{equation}
    N^{\text{D}} = \left( \sum_{k,j} | \mathcal{A}^{\text{D}}_{\mu,e;k,j} |^2 \right)^{-1}. \label{normD}
\end{equation}

The normalized density matrix that describes the detected $\nu_{\mu}$ is given by the trace over the initial electron state:
\begin{eqnarray}
    \hat{\rho}^{\text{D}}_{\nu_{\mu}} &=& \langle e^{-} | \hat{\rho}^{\text{D}} | e^{-} \rangle \nonumber \\ 
    &=&  N^{\text{D}} \sum_{j,k,k'} \mathcal{A}^{\text{D}}_{\mu,e;k,j} \mathcal{A}^{\text{D}*}_{\mu,e;k',j} | \nu_{k} \rangle \langle \nu_{k'} | \nonumber \\
    &=& N^{\text{D}} \sum_{j} |U_{ej}|^2 \sum_{k,k'} U_{\mu k}^{*} U_{\mu k'} \mathcal{M}^{\text{D}}_{k,j} \mathcal{M}^{\text{D}*}_{k',j} | \nu_{k} \rangle \langle \nu_{k'}|. \ \ \ \ \ \ \label{dmDnumu}
\end{eqnarray}
If reaction (\ref{imd}) is used in a neutrino oscillation experiment
with initial $\nu_{e}$'s produced by the decay of $\mu^{+}$
(for example, in a Neutrino Factory)
and described by the density matrix
(\ref{rho_nue}),
we can calculate an oscillation probability associated with the reaction by\footnote{See the similar treatment in Ref.~\cite{Szafron:2011zz}.
Note that the set
$\{ \hat{\rho}^{\text{D}}_{\nu_{\mu}}, \mathsf{I} - \hat{\rho}^{\text{D}}_{\nu_{\mu}} \}$
can be considered as a discrete unsharp positive operator-valued measure (POVM); see Ref.~\cite{Gudder:2007}.
}
\begin{equation}
\mathsf{P}_{e \to \mu}
=
\text{Tr} \left[ \hat{\rho}_{\nu_e}(T,\vec{L}) \, \hat{\rho}^{\text{D}}_{\nu_{\mu}} \right].
\label{Pemu}
\end{equation}

We omit the lengthy explicit expression of the probability
resulting from Eq.~(\ref{Pemu}),
that can be calculated straightforwardly, but note that it can be shown that $\mathsf{P}_{e \to \mu} \leq 1$.
We also emphasize that in the usual approximation in which
the differences of the neutrino masses are negligible in the production and detection processes
we have
$\mathcal{M}^{\text{P}}_{k,j} \simeq \mathcal{M}^{\text{P}}$
and
$\mathcal{M}^{\text{D}}_{k,j} \simeq \mathcal{M}^{\text{D}}$,
$\forall j,k$.
In this approximation we recover the standard expression in Eq.~(\ref{Pstd}) for the oscillation probability.
Also in the case of
$N_{\ell}$ light and $N_{d}$ heavy decoupled neutrinos
discussed in Section~\ref{sec:IIA} we obtain
a description in terms of
the flavor states in Eq.~(\ref{heavy1}).

On the other hand,
the case of
$N_{\ell}$ light neutrinos,
$N_{h}$ heavy neutrinos which are produced and detected incoherently,
and $N_{d}$ heavy decoupled neutrinos
is rather complicated.
One can derive the detection density matrix in analogy with
the production density matrix in Eq.~(\ref{inc1}).
Then, in analogy with the derivation of Eq.~(\ref{inc2})
we obtain the $\nu_{e}\to\nu_{\mu}$ experimental rate
\begin{align}
R_{e\mu}(L,E)
\propto
\null & \null
\left( \sum_{j \leq N_{\ell}} |U_{\mu j}|^2 \right)
\left( \sum_{k \leq N_{\ell}} |U_{e k}|^2 \right)
\nonumber
\\
\null & \null
\hspace{1cm}
\times
\Gamma_{\mu^{+}}^{0,0}(E)
\mathsf{P}_{\nu_{e}\to\nu_{\mu}}^{\text{eff}}(L,E)
\sigma_{\mu}^{0,0}(E)
\nonumber
\\
\null & \null
\hspace{-1cm}
+
\left( \sum_{k \leq N_{\ell}} |U_{e k}|^2 \right)
\nonumber
\\
\null & \null
\hspace{-0.5cm}
\times
\sum_{j=N_{\ell}+1}^{N_{\ell}+N_{h}} |U_{\mu j}|^2
\Gamma_{\mu^{+}}^{0,j}(E)
\mathsf{P}_{\nu_{e}\to\nu_{\mu}}^{\text{eff}}(L,E)
\sigma_{\mu}^{0,0}(E)
\nonumber
\\
\null & \null
\hspace{-1cm}
+
\left( \sum_{j \leq N_{\ell}} |U_{\mu j}|^2 \right)
\nonumber
\\
\null & \null
\hspace{-0.5cm}
\times
\sum_{k=N_{\ell}+1}^{N_{\ell}+N_{h}} |U_{e k}|^2
\Gamma_{\mu^{+}}^{0,0}(E)
\mathsf{P}_{\nu_{e}\to\nu_{\mu}}^{\text{eff}}(L,E)
\sigma_{\mu}^{0,k}(E)
\nonumber
\\
\null & \null
\hspace{-1cm}
+
\sum_{j,k \leq N_{\ell}+N_{h}} |U_{\mu j}|^2 |U_{e k}|^2
\nonumber
\\
\null & \null
\hspace{-0.5cm}
\times
\sum_{i=N_{\ell}+1}^{N_{\ell}+N_{h}}
|U_{e i}|^2
|U_{\mu i}|^2
\Gamma_{\mu^{+}}^{i,j}(E)
\sigma_{\mu}^{i,k}(E)
,
\label{inc3}
\end{align}
where
$\sigma_{\mu}^{i,k}(E)$
is the $\nu_{i} + e^{-} \to \mu^{-} + \nu_{k}$ cross section.
This equation shows that the experimental rate depends not only on the heavy neutrino masses
which constitute the detected neutrino,
but also on those which constitute the undetected neutrino in the production process
and the undetected final neutrino in the detection process.

\section{Neutrino-electron elastic scattering}\label{sec:IVb}

Neutrinos can also be detected with the neutrino-electron elastic scattering (ES) process
\begin{equation}
    \nu + e^{-} \to \nu + e^{-}. \label{ES}
\end{equation}
This is a more complicated case,
because it is not a pure charged-current interaction in which a leptonic flavor is selected.
However,
there is a flavor dependence, due to the fact that
$\nu_{\mu}$'s and $\nu_{\tau}$'s interact only through neutral currents,
whereas $\nu_{e}$'s interact through both charged and neutral currents.
For example, in water Cherenkov solar neutrino experiments
information on solar neutrino oscillations is obtained
by observing the ES reaction (\ref{ES}) induced by solar neutrinos,
taking into account that the cross section
$\sigma^{\text{ES}}_{\nu_{e}}(E)$
of $\nu_{e}$'s is about six times larger than the cross section
$\sigma^{\text{ES}}_{\nu_{\mu,\tau}}(E)$
of $\nu_{\mu}$'s and $\nu_{\tau}$'s.
In these experiments,
the rate of ES events in a detector is calculated as
\begin{align}
R^{\text{ES}}
=
\null & \null
N_{e} \int \text{d}E \, \phi_{\nu_{e}}(E)
\Big[
\mathsf{P}_{\nu_{e}\to\nu_{e}}(E) \sigma^{\text{ES}}_{\nu_{e}}(E)
\nonumber
\\
\null & \null
\hspace{1cm}
+
\sum_{\alpha=\mu,\tau}
\mathsf{P}_{\nu_{e}\to\nu_{\alpha}}(E)
\sigma^{\text{ES}}_{\nu_{\mu,\tau}}(E)
\Big]
,
\label{ESrate}
\end{align}
where
$N_{e}$ is the number of electrons in the detector,
$\phi_{\nu_{e}}(E)$ is the solar $\nu_{e}$ flux,
and
$\mathsf{P}_{\nu_{e}\to\nu_{\alpha}}(E)$
is the probability of $\nu_{e}\to\nu_{\alpha}$ oscillations
from the center of the Sun to the detector.
Note that the standard cross sections
$\sigma^{\text{ES}}_{\nu_{e}}(E)$
and
$\sigma^{\text{ES}}_{\nu_{\mu,\tau}}(E)$
are calculated neglecting the neutrino masses.

In the following we present a schematic calculation of
$R^{\text{ES}}$
which takes into account the neutrino masses in the interaction process
and we show that it reduces to the expression in Eq.~(\ref{ESrate})
only in the standard framework of mixing of three light neutrinos
and in its extensions with light sterile neutrinos.

It is possible in principle to define a density matrix
which describes the neutrino detected in the ES process
(\ref{ES})
following a method similar to that presented in Sec.~\ref{sec:IV},
but such a density matrix is not useful to obtain the rate $R^{\text{ES}}$,
where the oscillation probability and the cross section are not factorized.
Therefore, we calculate directly $R^{\text{ES}}$
considering a neutrino with energy $E$ coming from the Sun, which is described by the state
\begin{equation}
| \nu_{\text{S}}(E) \rangle
=
\sum_{k}
\mathsf{A}_{\nu_{e}\to\nu_{k}}(E) | \nu_{k}(E) \rangle
,
\label{nusun}
\end{equation}
where $\mathsf{A}_{\nu_{e}\to\nu_{k}}$ is the amplitude of $\nu_{e}\to\nu_{k}$
transitions from the center of the Sun to the detector.
In practice,
in the standard framework of three-neutrino mixing,
the established values of the neutrino masses and mixing imply that
solar neutrinos arrive at Earth as effectively incoherent sums
of mass eigenstates
\cite{Dighe:1999id,Smirnov:2016xzf}.
Therefore, the measurable oscillation probability is obtained by omitting the
interference terms between different massive neutrino contributions.
However, since we consider the general theory,
which in principle allows the possibility of vacuum oscillations between the Sun and the Earth
due to very small mass splittings
(see, for example, Ref.~\cite{Giunti:2007ry}),
we consider the description in Eq.~(\ref{nusun})
with the additional prescription of omitting, when needed, the interference terms
between incoherent massive neutrino contributions
in the calculation of the oscillation probability.

Since the final neutrino
in the detection process (\ref{ES})
is a superposition of orthogonal massive neutrinos,
the cross section is,
similarly to that of process (\ref{imd}),
the incoherent sum of the cross sections of the processes
\begin{equation}
    \nu + e^{-} \to \nu_{j} + e^{-}. \label{ESj}
\end{equation}
The rate of ES events in a detector is given by
\begin{equation}
R^{\text{ES}}
=
N_{e} \int \text{d}E \ \phi_{\nu_{e}}(E) \sigma_{\text{S}}(E)
,
\label{ESrate2}
\end{equation}
with
\begin{equation}
\sigma_{\text{S}}(E)
=
\int \text{dPS}
\sum_{j}
\left|
\langle \nu_{j}, e^{-} | (\mathsf{S} - \mathsf{I}) | \nu_{\text{S}}(E), e^{-} \rangle
\right|^2
,
\label{sigmaS}
\end{equation}
where the integration over $\text{dPS}$ represents schematically the integration over the phase space.
In this case, we must consider an expansion of the $\mathsf{S}$ matrix which contains,
in addition to the charged-current weak interactions already considered in Eq.~(\ref{S_CC}),
also neutral-current interactions:
\begin{equation}
  \mathsf{S}
  \approx
  \mathsf{I}
  - i \, \frac{G_{\text{F}}}{\sqrt{2}} \int \text{d}^{4}x
  \left[
  \hat{j}_{\text{CC}\rho}^{\dagger}(x) \hat{j}_{\text{CC}}^\rho(x)
  +
  \hat{j}_{\text{NC}\rho}(x) \hat{j}_{\text{NC}}^\rho(x)
  \right].
  \label{S_CC+NC}
\end{equation}
Considering the possible existence of sterile (light or heavy) neutrinos
beyond the standard framework of three-neutrino mixing,
the weak neutral current is given by
\begin{align}
\hat{j}_{\text{NC}}^\rho(x)
=
\null & \null
\frac{1}{2}
\sum_{j,k}
\sum_{\alpha=e,\mu,\tau}
U_{\alpha j}^{*}
U_{\alpha k}
\hat{\overline{\nu}}_{j}(x) \gamma^{\rho} \left( 1 - \gamma^{5} \right) \hat{\nu}_{k}(x)
\nonumber
\\
\null & \null
+
\hat{\overline{e}}(x) \gamma^{\rho} \left( g_{V}^{e} - \gamma^{5} g_{A}^{e} \right) \hat{e}(x)
,
\label{jNC}
\end{align}
where
$g_{V}^{e} = - 1/2 + 2 \sin^2 \vartheta_{W}$
and
$g_{A}^{e} = - 1/2$,
and $\vartheta_{W}$ is the weak mixing angle.
Note the possible existence of flavor-changing neutral currents
due to the failure \cite{Schechter:1980gr} of the GIM mechanism \cite{Glashow:1970gm}
in the presence of sterile neutrinos
(since in this case $\sum_{\alpha=e,\mu,\tau} U_{\alpha j}^{*} U_{\alpha k} \neq \delta_{jk}$).
Then, $\sigma_{\text{S}}(E)$ is given by
\begin{align}
\sigma_{\text{S}}(E)
=
\null & \null
\int \text{dPS}
\,
\sum_{j}
\Big|
\sum_{k}
\mathsf{A}_{\nu_{e}\to\nu_{k}}(E)
\nonumber
\\
\null & \null
\times
\Big\{
U_{ej}^{*}
U_{ek}
\left[
\mathcal{M}^{\text{CC}}_{j,k}(E)
+
\mathcal{M}^{\text{NC}}_{j,k}(E)
\right]
\nonumber
\\
\null & \null
\hspace{0.5cm}
+
\sum_{\alpha=\mu,\tau}
U_{\alpha j}^{*}
U_{\alpha k}
\mathcal{M}^{\text{NC}}_{j,k}(E)
\Big\}
\Big|^2
,
\label{sigmaS2}
\end{align}
with the charged-current (CC) and neutral-current (NC) matrix elements
\begin{align}
\mathcal{M}^{\text{CC}}_{j,k}(E)
=
- i \,
\null & \null
\frac{G_{\text{F}}}{\sqrt{2}}
\int \text{d}^{4}x
\langle \nu_{j}, e^{-} |
\hat{\overline{e}}(x) \gamma^{\rho} \left( 1 - \gamma^{5} \right) \hat{\nu}_{k}(x)
\nonumber
\\
\null & \null
\times
\hat{\overline{\nu}}_{j}(x) \gamma_{\rho} \left( 1 - \gamma^{5} \right) \hat{e}(x)
| \nu_{k} , e^{-} \rangle
,
\label{MCC}
\\
\mathcal{M}^{\text{NC}}_{j,k}(E)
=
- i \,
\null & \null
\frac{G_{\text{F}}}{\sqrt{2}}
\int \text{d}^{4}x
\langle \nu_{j}, e^{-} |
\hat{\overline{e}}(x) \gamma^{\rho} \left( g_{V}^{e} - \gamma^{5} g_{A}^{e} \right) \hat{e}(x)
\nonumber
\\
\null & \null
\times
\hat{\overline{\nu}}_{j}(x) \gamma_{\rho} \left( 1 - \gamma^{5} \right) \hat{\nu}_{k}(x)
| \nu_{k} , e^{-} \rangle
.
\label{MNC}
\end{align}
We omit the expression of $\sigma_{\text{S}}(E)$
obtained from the evaluation of the squared modulus
in the general expression in Eq.~(\ref{sigmaS2})
because it does not yield any simplification
and the resulting expression is rather cumbersome.
Let us only note that in general there are interference terms
between
the $\nu_{e}$ terms
$
U_{ej}^{*}
U_{ek}
\left[
\mathcal{M}^{\text{CC}}_{j,k}(E)
+
\mathcal{M}^{\text{NC}}_{j,k}(E)
\right]
$
and the $\nu_{\mu,\tau}$ terms
$
\sum_{\alpha=\mu,\tau}
U_{\alpha j}^{*}
U_{\alpha k}
\mathcal{M}^{\text{NC}}_{j,k}(E)
$
that do not allow a separation of the
corresponding cross sections as in the standard expression (\ref{ESrate})
for the ES event rate.
These interference terms disappear in
the standard framework of mixing of three light neutrinos
and in its extensions with light sterile neutrinos,
where we can approximate
$\mathcal{M}^{\text{CC}}_{j,k}(E) \simeq \mathcal{M}^{\text{CC}}_{0,0}(E)$
and
$\mathcal{M}^{\text{NC}}_{j,k}(E) \simeq \mathcal{M}^{\text{NC}}_{0,0}(E)$
$\forall k,j$.
In this case, taking into account the unitarity relation
\begin{equation}
\sum_{j} U_{\alpha j}^{*} U_{\beta j} = \delta_{\alpha\beta}
,
\label{unitarity}
\end{equation}
we obtain
\begin{align}
\sigma_{\text{S}}(E)
\simeq
\int \text{dPS}
\null & \null
\Big[
\mathsf{P}_{\nu_{e}\to\nu_{e}}(E)
\left|
\mathcal{M}^{\text{CC}}_{0,0}(E)
+
\mathcal{M}^{\text{NC}}_{0,0}(E)
\right|^2
\nonumber
\\
\null & \null
+
\sum_{\alpha=\mu,\tau}
\mathsf{P}_{\nu_{e}\to\nu_{\alpha}}(E)
|\mathcal{M}^{\text{NC}}_{0,0}(E)|^2
\Big]
,
\label{sigmaS4}
\end{align}
where
$\mathsf{P}_{\nu_{e}\to\nu_{\alpha}} = | \sum_{k} \mathsf{A}_{\nu_{e}\to\nu_{k}} U_{\alpha k} |^2$.
When the squared moduli of the interaction matrix elements are integrated over the phase space, they give the cross sections in Eq.~(\ref{ESrate}):
\begin{equation}
\sigma_{\text{S}}(E)
\simeq
\mathsf{P}_{\nu_{e}\to\nu_{e}}(E)
\sigma^{\text{ES}}_{\nu_{e}}(E)
+
\sum_{\alpha=\mu,\tau}
\mathsf{P}_{\nu_{e}\to\nu_{\alpha}}(E)
\sigma^{\text{ES}}_{\nu_{\mu,\tau}}(E)
,
\label{sigmaS5}
\end{equation}
with the standard cross sections
$\sigma^{\text{ES}}_{\nu_{e}}(E)$
and
$\sigma^{\text{ES}}_{\nu_{\mu,\tau}}(E)$
calculated neglecting the neutrino masses.
Hence,
the expression (\ref{ESrate}) used to calculate the
rate of ES events in water Cherenkov solar neutrino experiments
is correct in the standard framework of mixing of three light neutrinos
and in its extensions with light sterile neutrinos.
Note that
the flavor changing neutral currents
present in the case of light sterile neutrinos
do not give any observable effect,
because the flavor of the final neutrino in the
elastic scattering process (\ref{ES}) is not observed.

As remarked after Eq.~(\ref{nusun}),
we described the neutrinos coming from the Sun
as coherent superpositions of massive neutrinos.
If solar neutrinos arrive at Earth as incoherent sums of the mass eigenstates
because of the separation of the corresponding wave packets,
the rate of ES events is obtained by summing incoherently
the different massive neutrino contributions.
This is equivalent to neglecting the interference terms
in the evaluation of the squared modulus in Eq.~(\ref{sigmaS2}),
but Eq.~(\ref{sigmaS4})
is obtained anyhow in the standard framework of mixing of three light neutrinos
and in its extensions with light sterile neutrinos.
The incoherence must be taken into account in the
calculation of the transition probabilities
$\mathsf{P}_{\nu_{e}\to\nu_{\alpha}}$,
that in the coherent case are given by
\begin{equation}
\mathsf{P}_{\nu_{e}\to\nu_{\alpha}}^{\text{coh}}
=
\left| \sum_{k} \mathsf{A}_{\nu_{e}\to\nu_{k}} U_{\alpha k} \right|^2
,
\label{Pcoh}
\end{equation}
whereas in the incoherent case they are given by
\begin{equation}
\mathsf{P}_{\nu_{e}\to\nu_{\alpha}}^{\text{inc}}
=
\sum_{k} |\mathsf{A}_{\nu_{e}\to\nu_{k}}|^2 |U_{\alpha k}|^2
,
\label{Pinc}
\end{equation}

Let us now consider the case of
$N_{\ell}$ light and $N_{d}$ heavy decoupled neutrinos,
in which
$k \leq N_{\ell}$
and
$\mathcal{M}^{\text{CC}}_{j,k}(E) = \mathcal{M}^{\text{NC}}_{j,k}(E) = 0$
for $j>N_{\ell}$,
whereas
$\mathcal{M}^{\text{CC}}_{j,k}(E) \simeq \mathcal{M}^{\text{CC}}_{0,0}(E)$
and
$\mathcal{M}^{\text{NC}}_{j,k}(E) \simeq \mathcal{M}^{\text{NC}}_{0,0}(E)$
for
$j \leq N_{\ell}$.
In this case, the expression of $\sigma_{\text{S}}(E)$
is different in the coherent and incoherent descriptions
of the neutrinos coming from the Sun.
Considering, for simplicity,
only the realistic incoherent case,
we obtain
\begin{align}
\null & \null
\sigma_{\text{S}}(E)
\simeq
\int \text{dPS}
\Bigg\{
\left|
\mathcal{M}^{\text{CC}}_{0,0}(E)
+
\mathcal{M}^{\text{NC}}_{0,0}(E)
\right|^2
\mathsf{P}_{\nu_{e}\to\nu_{e}}^{\text{inc,eff}}(E)
\nonumber
\\
\null & \null
\hspace{3cm}
\times
\sum_{j \leq N_{\ell}} |U_{e j}|^2
\nonumber
\\
\null & \null
+
|\mathcal{M}^{\text{NC}}_{0,0}(E)|^2
\sum_{k \leq N_{\ell}}
|\mathsf{A}_{\nu_{e}\to\nu_{k}}(E)|^2
\sum_{j \leq N_{\ell}}
\left|
\sum_{\alpha=\mu,\tau}
U_{\alpha j}^{*}
U_{\alpha k}
\right|^2
\nonumber
\\
\null & \null
+
2
\left[
\mathcal{M}^{\text{CC}}_{0,0}(E)
+
\mathcal{M}^{\text{NC}}_{0,0}(E)
\right]
\mathcal{M}^{\text{NC}*}_{0,0}(E)
\sum_{k \leq N_{\ell}}
|\mathsf{A}_{\nu_{e}\to\nu_{k}}(E)|^2
\nonumber
\\
\null & \null
\hspace{2cm}
\times
\operatorname{Re}
\sum_{j \leq N_{\ell}}
U_{e j}^{*}
U_{e k}
\sum_{\alpha=\mu,\tau}
U_{\alpha j}
U_{\alpha k}^{*}
\Bigg\}
,
\label{sigmaS6}
\end{align}
where
\begin{equation}
\mathsf{P}_{\nu_{e}\to\nu_{e}}^{\text{inc,eff}}
=
\sum_{k \leq N_{\ell}} |\mathsf{A}_{\nu_{e}\to\nu_{k}}|^2 |U_{e k}|^2
,
\label{Pinceff}
\end{equation}
Integrating over the phase space,
we obtain
\begin{align}
\null & \null
\sigma_{\text{S}}(E)
\simeq
\sigma^{\text{ES}}_{\nu_{e}}(E)
\mathsf{P}_{\nu_{e}\to\nu_{e}}^{\text{inc,eff}}(E)
\sum_{j \leq N_{\ell}} |U_{e j}|^2
\nonumber
\\
\null & \null
\hspace{0.7cm}
+
\sigma^{\text{ES}}_{\nu_{\mu,\tau}}(E)
\sum_{k \leq N_{\ell}}
|\mathsf{A}_{\nu_{e}\to\nu_{k}}(E)|^2
\sum_{j \leq N_{\ell}}
\left|
\sum_{\alpha=\mu,\tau}
U_{\alpha j}^{*}
U_{\alpha k}
\right|^2
\nonumber
\\
\null & \null
\hspace{0.7cm}
+
\sigma^{\text{ES}}_{\text{int}}(E)
\sum_{k \leq N_{\ell}}
|\mathsf{A}_{\nu_{e}\to\nu_{k}}(E)|^2
\nonumber
\\
\null & \null
\hspace{2cm}
\times
\operatorname{Re}
\sum_{j \leq N_{\ell}}
U_{e j}^{*}
U_{e k}
\sum_{\alpha=\mu,\tau}
U_{\alpha j}
U_{\alpha k}^{*}
.
\label{sigmaS7}
\end{align}
This expression is rather complicated
and quite different from the standard one in Eq.~(\ref{sigmaS5}),
especially for the contribution of the new cross section
$\sigma^{\text{ES}}_{\text{int}}(E)$
obtained from the interference of the
$\nu_{e}$
and
$\nu_{\mu,\tau}$
interaction amplitudes.
The presence of this interference cross section is due to the
non-orthogonality of the
$\nu_{e}$
and
$\nu_{\mu,\tau}$
states discussed in Section~\ref{sec:IIA}.
We can write $\sigma^{\text{ES}}_{\text{int}}(E)$ as
\begin{equation}
\sigma^{\text{ES}}_{\text{int}}(E)
=
\sigma^{\text{ES}}_{\nu_{e}}(E)
+
\sigma^{\text{ES}}_{\nu_{\mu,\tau}}(E)
-
\sigma^{\text{ES,CC}}_{\nu_{e}}(E)
,
\label{sigmaint1}
\end{equation}
where
$\sigma^{\text{ES,CC}}_{\nu_{e}}(E)$
is the cross section of $\nu_{e}$ due to charged-current interactions only,
which is obtained from the standard expression of
$\sigma^{\text{ES}}_{\nu_{e}}(E)$
(see, for example, Ref.~\cite{Giunti:2007ry})
by replacing the real values of 
$g_{V}^{e}$
and
$g_{A}^{e}$
with
$g_{V}^{e} = g_{A}^{e} = 0$.

In the more general case of
$N_{\ell}$ light neutrinos which are detected incoherently,
$N_{h}$ heavy neutrinos which are produced and detected incoherently,
and $N_{d}$ heavy decoupled neutrinos,
$\sigma_{\text{S}}(E)$
is obtained by adding to Eq.~(\ref{sigmaS7})
\begin{align}
\Delta\sigma_{\text{S}}(E)
=
\null & \null
\int \text{dPS}
\Bigg\{
\sum_{j = N_{\ell}+1}^{N_{\ell}+N_{h}}
\sum_{k \leq N_{\ell}}
|\mathsf{A}_{\nu_{e}\to\nu_{k}}(E)|^2
\Lambda_{j,k}(E)
\nonumber
\\
\null & \null
+
\sum_{j \leq N_{\ell}+N_{h}}
\sum_{k = N_{\ell}+1}^{N_{\ell}+N_{h}}
|U_{ek}|^2
\Lambda_{j,k}(E)
\Bigg\}
,
\label{sigmaS8}
\end{align}
with
\begin{align}
\Lambda_{j,k}(E)
=
\null & \null
|U_{ej}|^2
|U_{ek}|^2
\left|
\mathcal{M}^{\text{CC}}_{j,k}(E)
+
\mathcal{M}^{\text{NC}}_{j,k}(E)
\right|^2
\nonumber
\\
\null & \null
+
\sum_{\alpha,\beta=\mu,\tau}
U_{\alpha j}^{*}
U_{\alpha k}
U_{\beta j}
U_{\beta k}^{*}
|\mathcal{M}^{\text{NC}}_{j,k}(E)|^2
\nonumber
\\
\null & \null
+
2
\operatorname{Re}
U_{ej}^{*}
U_{ek}
\sum_{\alpha=\mu,\tau}
U_{\alpha j}
U_{\alpha k}^{*}
\nonumber
\\
\null & \null
\hspace{0.3cm}
\times
\left[
\mathcal{M}^{\text{CC}}_{j,k}(E)
+
\mathcal{M}^{\text{NC}}_{j,k}(E)
\right]
\mathcal{M}^{\text{NC}*}_{j,k}(E)
.
\label{sigmaS9}
\end{align}
In Eq.~(\ref{sigmaS8})
we took into account that
for the $N_{h}$ heavy neutrinos which are produced and detected incoherently
$|\mathsf{A}_{\nu_{e}\to\nu_{k}}|^2 = |U_{ek}|^2$
because their masses are much larger that the matter potential
that can change the effective mixing of light neutrinos in the Sun
with respect to that in vacuum.


\section{Conclusion}\label{sec:V}

Neutrino oscillations is one of the most interesting
phenomena in modern fundamental physics.
It was proposed about 60 years ago
\cite{Pontecorvo:1957cp,Pontecorvo:1957qd,Maki:1962mu},
and it has been observed about 20 years ago
\cite{Fukuda:1998mi,Ahmad:2002jz}.
Its standard theory is well known
(see, for example, Refs.~\cite{Giunti:2007ry,Bilenky:2018hbz,Patrignani:2016xqp}),
but it is also well known that it is an approximation and several subtle issues
have been discussed in the literature
(see, for example, the recent discussions in Refs.~\cite{Akhmedov:2010ua,Akhmedov:2012uu,Akhmedov:2017xxm}).
A particular subtle problem is the description of flavor neutrinos
\cite{Giunti:1991cb,Bilenky:2001yh,Giunti:2002xg,Giunti:2006fr}.

In this paper we discussed how to describe flavor neutrinos
produced or detected
in processes which involve more than one neutrino.
We have shown that in these cases flavor neutrinos cannot be described by pure states,
but require a density matrix description.
The density matrices can be approximated with density matrices of pure states
only when the differences of the neutrino masses are neglected
in the interaction process.
In this approximation,
the pure states are the standard flavor states and one recovers the standard expression
for the neutrino oscillation probability.

We discussed also the effects of mixing of the three standard light neutrinos
with heavy neutrinos which can be either decoupled
because their masses are much larger than the
maximum neutrino energy in the neutrino production process
or are produced and detected incoherently.
We have shown that in the case of only decoupled heavy neutrinos
the density matrix description
reduces to a description in terms of flavor states,
which however have the nonstandard features discussed in Section~\ref{sec:IIA}.
On the other hand,
in the presence of heavy neutrinos which are not decoupled
the density matrix description is non-reducible.
In a neutrino oscillation experiment
with production and detection processes involving multiple neutrinos
the experimental rate depends not only on the heavy neutrino masses
which constitute the detected neutrino,
but also on those which constitute the undetected neutrinos.

We also discussed
the more complicated case of neutrino-electron elastic scattering,
in which the flavors of the initial and a final neutrino
are not determined,
but there is a flavor dependence due to the different contributions of
charged-current and neutral-current interactions.
In this case it is not useful to define a density matrix
which describes the detected neutrino,
because the oscillation probability and the cross section are not factorized
in the detection rate.
As an example, we calculated the rate of neutrino-electron elastic scattering events
in a solar neutrino experiment
and we have shown that the usual expression in which the rate is given by the sum of the
$\nu_{e}$
and
$\nu_{\mu,\tau}$
contributions is obtained
only in the standard framework of mixing of three light neutrinos
and in its extensions with light sterile neutrinos.

Let us finally note that,
although for simplicity we considered as examples
processes in which there are only two neutrinos,
the formalism can be extended in a straightforward way
to the more complicated case of interactions involving more than two neutrinos.


\acknowledgments

G.~C. was fully supported by S\~ao Paulo Research Foundation (FAPESP) under grant 2016/08025-0.


\end{document}